\documentclass[aps,prl,reprint,bibnotes,groupedaddress,nofootinbib]{revtex4-2}
\usepackage{graphicx,amsmath}
\usepackage{xcolor}
\usepackage{tablefootnote}

\begin{document}


\title{Lattice Light Shift Evaluations In a Dual-Ensemble Yb Optical Lattice Clock}



\renewcommand{\andname}{\ignorespaces}

\author{Tobias Bothwell\textsuperscript{1,*}, Benjamin D. Hunt\textsuperscript{1,2,*}, Jacob L. Siegel\textsuperscript{1,2,*}, Youssef S. Hassan\textsuperscript{1,2},
	Tanner Grogan\textsuperscript{1,2}, Takumi Kobayashi\textsuperscript{1,4}, Kurt Gibble\textsuperscript{1,5}, Sergey G. Porsev\textsuperscript{6}, Marianna S. Safronova\textsuperscript{6}, Roger C. Brown\textsuperscript{1}, Kyle Beloy\textsuperscript{1}, Andrew D. Ludlow\textsuperscript{1,2,3,$\dagger$}}

\affiliation{\textsuperscript{1}National Institute of Standards and Technology, 325 Broadway, Boulder, Colorado 80305, USA}
\affiliation{\textsuperscript{2}University of Colorado, Department of Physics, Boulder, Colorado 80309, USA}
\affiliation{\textsuperscript{3}Electrical, Computer \& Energy Engineering, University of Colorado, Boulder, Colorado, USA}
\affiliation{\textsuperscript{4}National Metrology Institute of Japan (NMIJ), National Institute of Advanced Industrial Science and Technology (AIST), 1-1-1 Umezono, Tsukuba, Ibaraki 305-8563, Japan}
\affiliation{\textsuperscript{5}Department of Physics, The Pennsylvania State University, University Park, Pennsylvania 16802, USA}
\affiliation{\textsuperscript{6}Department of Physics and Astronomy, University of Delaware, Newark, Delaware 19716, USA}
\def\thefootnote{*}\footnotetext{These authors contributed equally to this Letter.}\def\thefootnote{\arabic{footnote}}
\def\thefootnote{$\dagger$}\footnotetext{Contact author: andrew.ludlow@nist.gov}\def\thefootnote{\arabic{footnote}}
\date{\today}

\begin{abstract}

In state-of-the-art optical lattice clocks, beyond-electric-dipole polarizability terms lead to a break-down of magic wavelength trapping. 
In this Letter, we report a novel approach to evaluate lattice light shifts, specifically addressing recent discrepancies in the atomic  multipolarizability term between experimental techniques and theoretical calculations. We combine imaging and multi-ensemble techniques to evaluate lattice light shift atomic coefficients, leveraging comparisons in a dual-ensemble lattice clock to rapidly evaluate differential frequency shifts. Further, we demonstrate application of a running wave field to probe both the multipolarizability and hyperpolarizability coefficients, establishing a new technique for future lattice light shift evaluations.

\end{abstract}


\maketitle


Optical lattice clocks (OLCs) are among the most accurate~\cite{ushijima2015cryogenic,mcgrew2018atomic,bothwell2019jila,aeppli2024clock} and precise~\cite{schioppo2017ultrastable,oelker2019demonstration,zheng2022differential,bothwell2022resolving} 
sensors ever created by humankind, positioning them as strong candidates for the redefinition of the SI second~\cite{dimarcq2024roadmap}. Modern clock performance further supports studies of fundamental physics, from searches for dark matter~\cite{derevianko2014hunting,kennedy2020precision} to tests of general relativity~\cite{takamoto2020test,zheng2023lab}. In parallel, emerging transportable OLCs promise to revolutionize relativistic geodesy, mapping Earth's geoid to new levels~\cite{mehlstaubler2018atomic}.

Central to OLC performance is the trapping of ultracold atoms at the so-called magic wavelength (or frequency)~\cite{takamoto2003spectroscopy,ye2008quantum}, where the differential dynamic polarizability between clock electronic states vanishes. The resulting differential light shift is fundamental to OLCs and is an accuracy-limiting systematic effect~\cite{mcgrew2018atomic,aeppli2024clock}. Higher-order perturbations from magic wavelength trapping, such as magnetic-dipole and electric-quadrupole terms (so-called multipolarizability)~\cite{taichenachev2008frequency}, produce nontrival couplings between the resulting light shifts and the motional states of the atomic sample, challenging the efficacy of magic wavelength trapping. Careful characterization of these shifts is ongoing. Multiple
experimental evaluations of these higher-multipolar corrections in $^{87}$Sr~\cite{ushijima2018operational,dorscher2023experimental,kim2023evaluation}, combined with recent theoretical development~\cite{WuShiNi23,porsev2023contribution}, have resolved disagreement of both the sign and magnitude of the multipolarizabilty coefficient. In $^{171}$Yb, disagreement remains between a single experimental result~\cite{nemitz2019modeling} and theoretical calculations~\cite{KatOvsMar15,OvsMarPal16,brown2017hyperpolarizability}.

Simultaneously, recent efforts have demonstrated how imaging techniques combined with multi-ensemble operation may be used to enhance the measurement capabilities of OLCs~\cite{marti2018imaging}. For example, differential measurements made by synchronous comparison between multiple optical clocks~\cite{oelker2019demonstration,schioppo2017ultrastable} or within a single clock system~\cite{zheng2022differential} reject common mode laser noise, realizing an effective decoherence-free subspace~\cite{marti2018imaging,manovitz2019precision}. Such techniques in 1D OLCs have demonstrated remarkable progress, observing the gravitational redshift at the millimeter scale~\cite{zheng2023lab} and utilizing multi-apparatus operation for extended coherence times~\cite{kim2023AlionZDT, schioppo2017ultrastable}. 

\begin{figure}[h!]
	\includegraphics[trim={0 3.3cm 0 3.1cm},clip,width=0.48\textwidth]{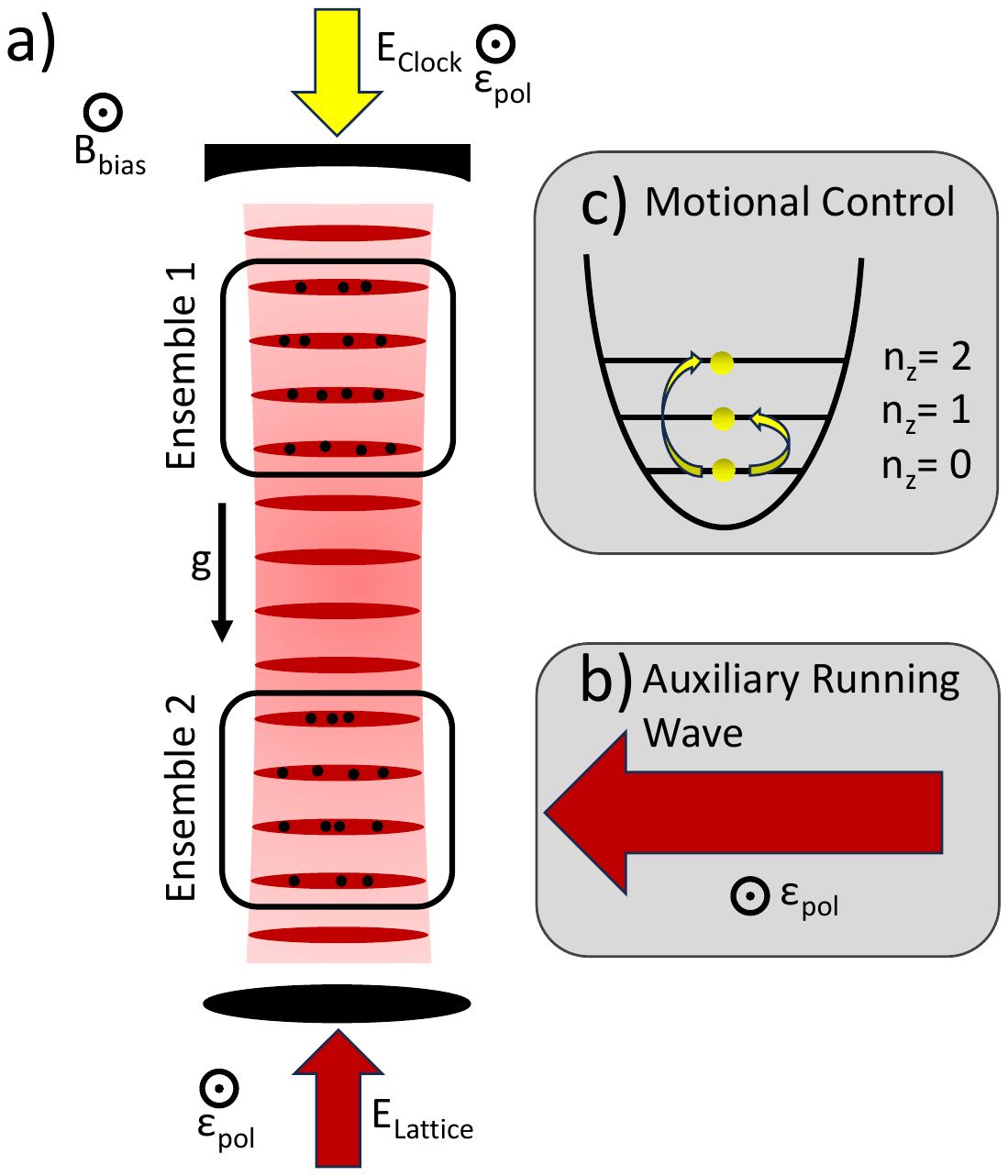}
	\caption{\label{1}(a) Schematic of a dual-ensemble 1D Yb OLC (not to scale). The 759 nm lattice is formed via retro-reflection of a single beam and clock light is introduced through the mirror used for reflection of the 759 nm beam. The directions of the polarization, magnetic field, and gravity orientation are indicated. (b) A 759 nm transverse running wave may be introduced to ensemble 2, allowing evaluation of the running wave magic wavelength and hyperpolarizability via differential comparisons. (c) The longitudinal motional state of the atoms in each ensemble may be manipulated separately, providing enhanced, differential sensitivity to higher-order light shift terms.  }
	\label{figure_1}
\end{figure}

In this Letter we demonstrate application of emerging multi-ensemble techniques to a full differential polarizability evaluation in an Yb OLC. Our experimental apparatus, described in previous publications~\cite{mcgrew2018atomic, chen2024clock},  is a standard OLC utilizing a vertical retro-reflected 1D magic wavelength optical lattice at 759 nm. Here, we employ a recently demonstrated `ratchet loading' technique~\cite{hassan2024ratchet}. We load two spatially separated ensembles using a combination of magnetic field control during MOT operation and shelving to the metastable clock state (see Fig.~\ref{figure_1}). We then employ clock-mediated Sisyphus cooling~\cite{chen2024clock} to achieve radial temperatures of $\sim 600$ nK and sideband cooling to prepare atoms in the ground longitudinal band, providing a more uniform sampling of the lattice antinodes. This dual-ensemble preparation forms the basis of the experiments reported in this Letter, allowing differential measurements between the ensembles. Details of the dual-ensemble preparation are given in the Supplemental Material~\cite{SMcite}.

Near the magic wavelength, the lattice light shift $\delta \nu_{LS}$ can be written as a function of trap depth $U$, detuning $\delta_L$ of lattice frequency $\nu_L$ from the electric dipole ($E1$) magic frequency $\nu_{E1}$ ($\delta_L=\nu_L-\nu_{E1}$), radial temperature $T_r$, and longitudinal vibrational state $n_z$. For simplicity we follow Ref.~\cite{ushijima2018operational}, adopting a light shift model utilizing a harmonic basis (see Appendix A for a complementary treatment with a more general model). The lattice light shift is then given by 

\begin{equation}
	\begin{aligned}
		\frac{\delta \nu_{LS}(u,\delta_L,n_z)}{\nu_{c}} \approx& \bigg ( \frac{\partial \tilde{\alpha}_{E1}}{\partial \nu}\delta_L- \tilde{\alpha}_{M1E2} \bigg ) \bigg ( n_z +1/2 \bigg )u^{1/2} \\
		& -\bigg [ \frac{\partial \tilde{\alpha}_{E1}}{\partial \nu}\delta_L+ \frac{3}{2} \tilde{\beta} \bigg ( n_z^2 +n_z+\frac{1}{2}\bigg )\bigg ] u \\
		& +2 \tilde{\beta}\bigg (n_z +\frac{1}{2} \bigg ) u^{3/2} - \tilde{\beta}u^2,
	\end{aligned}
\label{equation_lattice_light_shift}
\end{equation}

\noindent
where we have divided the clock shift ($\delta \nu_{LS}$) by the clock frequency ($\nu_c$) and utilize normalized trap depths $u=U/E_R$.  $E_R = (h\nu_L)^2/2mc^2$ is the recoil energy and $c$ the speed of light, $m$ the atomic mass, and $h$ Planck's constant. The effects of transverse temperatures are captured via an effective depth $u^j \rightarrow (1+jk_BT_r/uE_R)^{-1}u^j$~\cite{ushijima2018operational,beloy2020modeling} where $j$ is the power series exponent for each term in Eq.~(\ref{equation_lattice_light_shift}). $k_B$ is the Boltzmann constant and trap depth is measured via sideband spectroscopy~\cite{blatt2009sideband}. All trap depths $u^j$ in the Letter implicitly assume this effective radial thermal averaging.

Complete lattice light shift evaluations require knowledge of $\nu_{E1}$ and the three differential atomic coefficients within Eq.~(\ref{equation_lattice_light_shift}). $\frac{\partial \tilde{\alpha}_{E1}}{\partial \nu}$ is the linear slope of the differential $E1$ polarizability between the ground ($^1$S$_0$) and excited ($^3$P$_0$) clock states arising from a Taylor expansion about $\nu_{E1}$. 
$ \tilde{\alpha}_{M1E2}$ and $\tilde{\beta}$ are the differential multipolarzability and hyperpolarizability, respectively. 
These coefficients are often evaluated via interleaved comparisons between two trap depths ($u$)~\cite{brown2017hyperpolarizability} or two motional states ($n_z$)~\cite{ushijima2018operational}. By operating over a broad range of trap depths, lattice frequencies, and motional states, individual polarizability terms can be disentangled and measured.
In many OLCs, however, practical limits of the realizable trap depths make such an evaluation daunting at the state-of-the-art level. 
Here, we overcome this limitation by supplementing the standard evaluation techniques with imaging and multi-ensemble operation.

 \begin{figure}
	\includegraphics[trim={0 .3cm 0 0.cm},clip,width=0.49\textwidth]{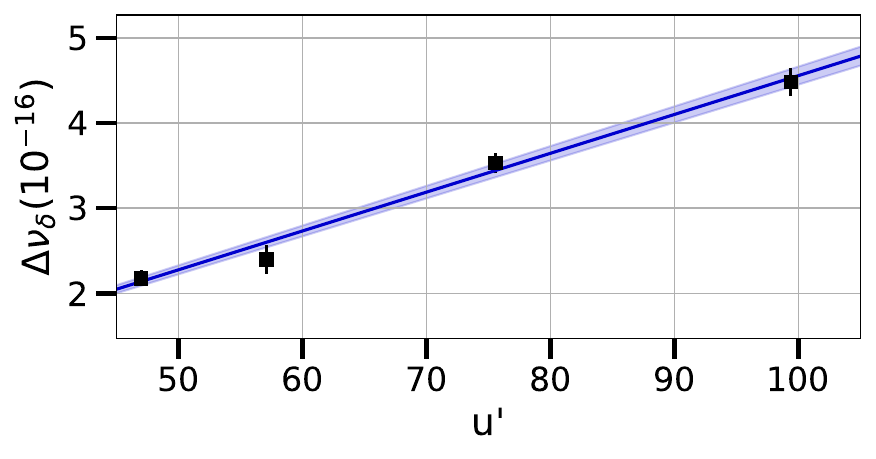}
	\caption{ $\frac{\partial \tilde{\alpha}_{E1}}{\partial \nu}$ is measured by temporally self-interleaving between lattice frequency detunings $\delta_2$ and $\delta_1$ for a single ensemble. The frequency shift, with error bars derived from the Allan deviation at half the total measurement time, is plotted versus $u'$. A linear fit, with  1-$\sigma$ error bars (shaded region), gives $\frac{\partial \tilde{\alpha}_{E1}}{\partial \nu}$ as the slope.  }
	\label{figure_2}
\end{figure}

\textit{Evaluation of $\partial \tilde{\alpha}_{E1}/\partial \nu$.}--The only terms in Eq.~(\ref{equation_lattice_light_shift}) that include $\partial \tilde{\alpha}_{E1}/\partial \nu$ are proportional to $\delta_L$. Therefore, self-interleaved measurements of the light shift at two lattice detunings $\delta_1$ and $\delta_2$ allow $\frac{\partial \tilde{\alpha}_{E1}}{\partial \nu}$ to be isolated. With the same initial preparation conditions, the frequency difference is  

\begin{equation}
	\begin{aligned}
		\frac{\Delta \nu_{\delta}(u,\delta_1,\delta_2,n_z)}{\nu_{c}} &=  -\frac{\partial \tilde{\alpha}_{E1}}{\partial \nu}\big ( \delta_2- \delta_1 \big )  u',
	\end{aligned}
	\label{equation_da_dv}
\end{equation}

\noindent
where we have introduced $u'= [u- (n_z+1/2)u^{1/2}]$. Critically, such a measurement is independent of $\tilde{\alpha}_{M1E2}$, $\tilde{\beta}$, and $\nu_{E1}$, while also benefiting from identical atom preparation to differentially reject cold collision shifts. As shown in Fig.~\ref{figure_2}, we perform these measurements at four trap depths with $\delta_{2}-\delta_1=-108.2(2)$ MHz and find $\frac{\partial \tilde{\alpha}_{E1}}{\partial \nu} = 4.2 (1) \times 10^{-20}$/MHz, in excellent agreement with previous measurements \cite{mcgrew2018atomic,kim2021absolute}. 

\textit{Evaluation of $\tilde{\beta}$.}--We now turn to the remaining atomic coefficients in Eq.~(\ref{equation_lattice_light_shift}). At the limited trap depths available in our apparatus ($<140$~$E_R$), evaluation of these shifts with standard interleaved measurements is challenging. Instead, we utilize imaging and dual-ensemble operation as shown in Fig.~\ref{figure_1}. Frequency comparisons between the two ensembles (found by converting differences in excitation probabilities to frequency via the known Rabi lineshape~\cite{SMcite}) are insensitive to laser frequency-noise, providing enhanced relative stability~\cite{marti2018imaging,bothwell2022resolving}. We regularly measure frequency instabilities of $\sim 4 \times 10^{-17}$ at 1 s for synchronous comparison between ensembles as compared with $\sim 3\times 10^{-16}$ for temporally self-interleaved measurements, allowing us to evaluate shifts nearly 50 times faster.

\begin{figure}
	\includegraphics[trim={0 3.4cm 0 3.5cm},clip,width=0.49\textwidth]{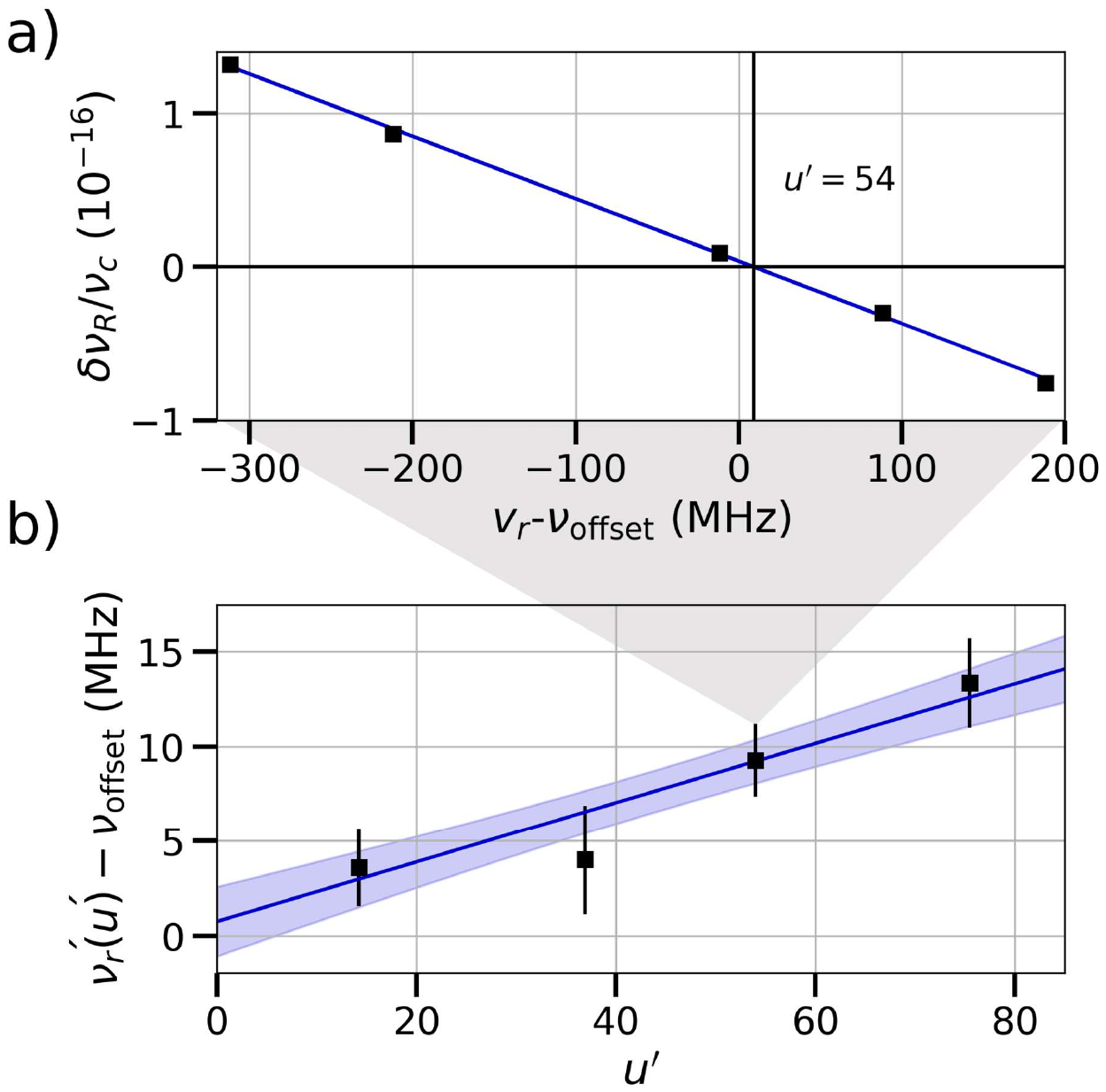}
	\caption{  Measurement of the running wave magic frequency, $\nu_{r}'(u')$. (a) The frequency shift arising from the addition of a running wave, measured synchronously, is plotted versus the running wave frequency, $\nu_r$. Error bars are smaller than the point size and $\nu_\mathrm{offset}=394~798~300$ MHz. We show the fit for a $u'=54$ standing wave contribution with black lines showing the fitted intercept for  $\nu_{r}'(u')$. (b) $\nu_{r}'(u')$ is plotted versus $u'$ for each of the four evaluated depths. The blue line and associated 1-$\sigma$ statistical uncertainty region show the fit to Eq.~(\ref{equation_running_experiment}).  }
	\label{figure_3}
\end{figure}

We apply an auxiliary running wave field to the second ensemble (Fig. 1(b)), near the magic frequency (but $>$MHz detuned from the standing wave laser frequency). For a running wave the $E1$ polarizability and multipolarizability terms simply add, in contrast to a standing wave where they are out of phase. The fractional frequency shift from the addition of an auxiliary running wave to the standing wave is

\begin{equation}
	\begin{aligned}
		  \frac{\delta \nu_{R}(u_r,u',\delta_r)}{\nu_c} & \approx- \bigg (  \frac{\partial \tilde{\alpha}_{E1}}{\partial \nu}\delta_{r} +\tilde{\alpha}_{M1E2} + \tilde{\beta}_d u' \bigg )u_r,  \\
	\end{aligned}
	\label{equation_running}
\end{equation}

\noindent
where $u_r$ is the running wave `depth', $u'$ the average standing wave depth experienced by the atoms (as introduced in Eq.~(\ref{equation_da_dv})), $\delta_{r}=\nu_r-\nu_{E1}$, and $\nu_r$ the running wave frequency (note that shifts of order $u_r^2$ and higher have been omitted here~\cite{SMcite}). Equation~(\ref{equation_running}) includes a shift term that is $\propto u' u_r$, arising from the dichromatic hyperpolarizability $\tilde{\beta}_d$~\cite{ovsiannikov2013multipole,beloyinprep2024}. For parallel linear lattice and running wave polarizations (Fig.~\ref{figure_1}), the dichromatic hyperpolarzability is related to the more familiar hyperpolarzability of Eq.~(\ref{equation_lattice_light_shift}) by $\tilde{\beta}_d =4 \tilde{\beta}$~\cite{beloyinprep2024}. This interference effect provides a new method to determine $\tilde{\beta}$ with minimal correlation to $\nu_{E1}$~\cite{brown2017hyperpolarizability}. Further, the use of synchronous dual-ensemble measurements facilitates its precise determination at shallow lattice depths. The auxiliary field has a $u'$-dependent frequency $\nu_r'(u')$ where $\delta \nu_{R}(u_r,u',\nu'_r-\nu_{E1})=0$, given by 

\begin{equation}
	\begin{aligned}
		\nu_r'(u') = \bigg ( \nu_{E1}-\frac{\tilde{\alpha}_{M1E2}}{ \frac{\partial \tilde{\alpha}_{E1}}{\partial \nu}} \bigg )- \frac{4 \tilde{\beta} u'}{\frac{\partial \tilde{\alpha}_{E1}}{\partial \nu}}.    \\
	\end{aligned}
	\label{equation_running_experiment}
\end{equation}

\noindent
$\nu_r'(u')$ is a linear function of $u'$ with a slope revealing $\tilde{\beta}$ and an offset $\nu_r'(0) =\nu_{E1}-\tilde{\alpha}_{M1E2}/ \frac{\partial \tilde{\alpha}_{E1}}{\partial \nu}$, directly relating $\nu_{E1}$ and $\tilde{\alpha}_{M1E2}$.

To experimentally evaluate $\nu_{r}'(u')$ we apply a running wave beam with a waist of $\approx 150$ $\mu$m to ensemble 2. We evaluate the ensemble-averaged depth to be $u_r \approx 10$, calibrated in-situ by dividing the slope of Fig.~3a by $-\frac{\partial \tilde{\alpha}_{E1}}{\partial \nu}$.  In this experiment, we do not apply Sisyphus cooling to lower the radial temperature, unlike all other measurements in this paper, as the addition of the running wave interferes with the optical access used for cooling.  At four different standing wave depths the running wave frequency is stepped over 500 MHz centered around the approximate location of $\nu_{r}'(u')$ (see Fig.~\ref{figure_3}). From these measurements a linear fit gives $\tilde{\beta} = -1.7(4)\times 10^{-21}$ and $\nu_r'(0)=394~798~300.4 (18)$ MHz. This value of $\tilde{\beta}$ falls between previous measurements using relatively deep optical lattices~\cite{brown2017hyperpolarizability,nemitz2019modeling} and is in good agreement with independent evaluations made via 2-photon resonances \cite{kobayashi2018uncertainty,Pizzocaro_2020}.

\textit{Evaluation of $\tilde{\alpha}_{M1E2}$ and $\nu_{E1}$.}--Returning to Fig.~\ref{figure_1}, we may prepare ensemble 1 in $n_z \approx 0$ and ensemble 2 in either $n_z \approx 1$ or $n_z \approx 2$~\cite{SMcite}. This allows differential comparisons between ensembles to be preferentially sensitive to the $\tilde{\alpha}_{M1E2}$ dominated $\sqrt{u}$ term of Eq. (1). The differential lattice light shift between samples with motional states $n_{1}$ and $n_{2}$ is given by

\begin{equation}
	\begin{aligned}
		\frac{\Delta \nu_{n}(u,\delta_r',n_1,n_2)}{\nu_{c}} \approx& \bigg ( \frac{\partial \tilde{\alpha}_{E1}}{\partial \nu}\delta_r'-2 \tilde{\alpha}_{M1E2} \bigg ) \big ( n_2-n_1  \big )u^{1/2} \\
		& -\frac{3}{2} \tilde{\beta} \bigg ( n_2^2 +n_2 -n_1^2-n_1 \bigg ) u  \\
		& +2 \tilde{\beta}\bigg (n_2-n_1 \bigg ) u^{3/2}
	\end{aligned}
	\label{equation_dnz}
\end{equation}

\noindent
with $\delta_r' = (\nu_L-\nu_r'(0))$. Note the elimination of $\nu_{E1}$ in Eq.~(\ref{equation_dnz}) by substitution of $\nu_r'(0)$ into Eq.~(\ref{equation_lattice_light_shift}). With the determinations of $\frac{\partial \tilde{\alpha}_{E1}}{\partial \nu}$, $\nu_{r}'(0)$, and $\tilde{\beta}$ in hand, this leaves only $\tilde{\alpha}_{M1E2}$ to evaluate.

 \begin{figure}
	\includegraphics[trim={0 .3cm 0 0.cm},clip,width=0.49\textwidth]{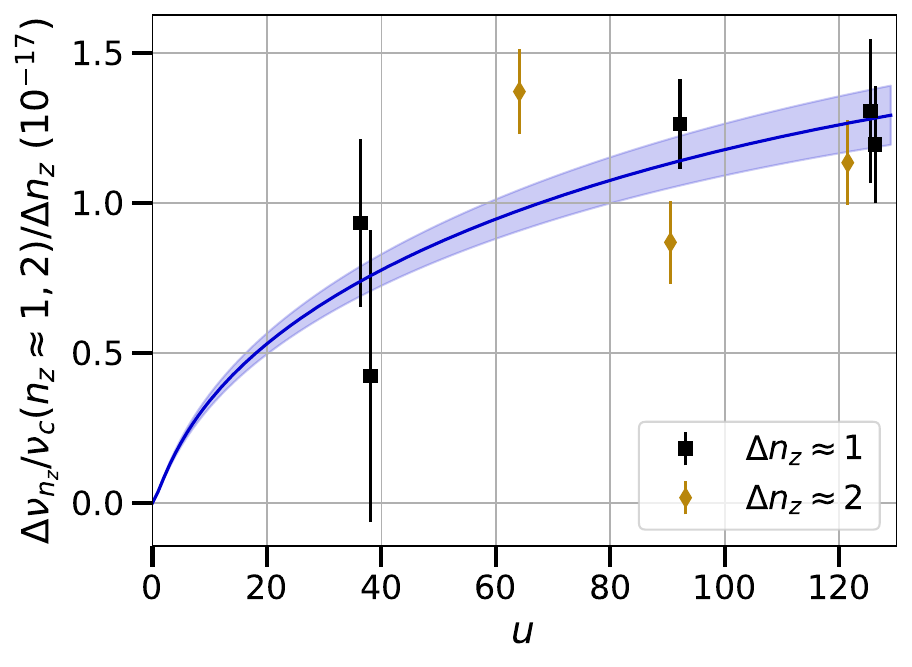}
	\caption{$\tilde{\alpha}_{M1E2}$ is measured via synchronous comparison of $n_z\approx0$ to $n_z\approx1$ (black squares) and to $n_z\approx2$ (gold diamonds).
		The fit to $\Delta \nu_{n_z}/\nu_c$, Eq.~(\ref{equation_dnz}), is shown in blue, with associated 1-$\sigma$ statistical uncertainty shaded. We plot the shifts normalized by $\Delta n_z=n_2-n_1$ to highlight the $\sqrt{u'}$ dependence predominantly arising from $\tilde{\alpha}_{M1E2}$. State-preparation errors resulted in $\Delta n_z\approx1 \rightarrow 0.8$ and $\Delta n_z\approx2 \rightarrow1.3$~\cite{SMcite}. As a result the plotted fit to Eq.~(\ref{equation_dnz}) is meant as a visual guide as it assumes perfect state preparation. A Monte-Carlo fit to Eq.~(\ref{equation_lattice_light_shift}) for each ensemble is required to fully account for $n_z$ and other experimental values, with the results in Table~\ref{table_summary}. A reduced chi-squared of 1.7 is found.}
	\label{figure_4}
\end{figure}

As shown in Fig.~\ref{figure_4}, we perform differential $n_z$ experiments at a variety of trap depths. The shift is shown normalized by the differential $n_z$ applied between ensembles, highlighting the $\sqrt{u}$ dependence (fit shown in blue). The radial temperatures are measured for each ensemble, and $n_z$-dependent cold collision corrections are applied~\cite{SMcite}. A Monte-Carlo method is used to propagate sources of uncertainty from both measured atomic coefficients and model inputs to the fit of each ensemble to Eq.~(\ref{equation_lattice_light_shift}). We find $\tilde{\alpha}_{M1E2} = -1.41(9) \times 10^{-18}$, in good agreement with a previous measurement at lower precision~\cite{nemitz2019modeling}. Finally, $\tilde{\alpha}_{M1E2}$ is substituted back into the definition of $\nu_r'(0)$, giving $\nu_{E1} = 394~798~266.9 (26)$ MHz. Table~\ref{table_summary} summarizes our experimental results.

\begin{table}[!h]
	\begin{center}
		\caption{Summary of experimental and theoretical values derived from this work. See Appendix A for a complementary Born-Oppenheimer + WKB treatment~\cite{beloy2020modeling}.}
		\label{table_summary}
		\begin{tabular}{c c} 
			\hline \hline  
			Coefficient & Value \\ [.5ex] 
			\hline  
			\rule{0pt}{1.\normalbaselineskip} $\frac{\partial \tilde{\alpha}_{E1}}{\partial \nu}$ ($10^{-20}$/MHz) & 4.2(1) \\ 
			
			$\tilde{\beta}$ ($10^{-21}$) & $-1.7(4)$ \\
			
			$\tilde{\alpha}_{M1E2}^{\text{Experiment}}$ $(10^{-18})$ & $-1.41(9)$ \\
			
			$\tilde{\alpha}_{M1E2}^{\text{Theory}}$ $(10^{-18})$& $-1.9(5)$ \\
			
			$\nu_{E1}$ (MHz) & $394~798~266.9 (26)$   \\
			
			$\nu'_r(0)$ (MHz) & $394~798~300.4 (18)$   \\ [1ex] 
			\hline \hline 
		\end{tabular}
	\end{center}
\end{table}

\textit{Theoretical predictions of $\tilde{\alpha}_{M1E2}$.}--It is now recognized that earlier calculations for Yb~\cite{KatOvsMar15,OvsMarPal16,brown2017hyperpolarizability}, Sr~\cite{ovsiannikov2013multipole,KatOvsMar15,OvsMarPal16,PorSafKoz18,WuTanShi19}, and other alkaline-earth(-like) systems~\cite{KatOvsMar15,OvsMarPal16,OvsMarMok17,WuTanShi20,PorSaf20,WuWanWan23} did not include the important diamagnetic contribution to the $M1$ polarizability at the magic wavelength. This resulted in a disagreement between theoretical and experimental results~\cite{ushijima2018operational,nemitz2019modeling,dorscher2023experimental,kim2023evaluation}, recently resolved in the case of Sr~\cite{WuShiNi23,porsev2023contribution}. The diamagnetic shift has been discussed extensively in the literature for the case of uniform dc magnetic fields (e.g., Refs.~\cite{Szm02,Kut03,ShiItaBol11}). In a nonrelativistic treatment, the diamagnetic shift appears at first order in perturbation theory and is proportional to the expectation value $\langle r^2\rangle$, where $r$ denotes the distance from the electron to the nucleus, a sum over all electrons is implied, and a total electronic angular momentum $J=0$ is assumed. In a relativistic treatment starting from the Dirac equation, the emergence of the diamagnetic shift is less conspicuous. It arises at second-order in perturbation theory, being attributed to negative-energy (positron) states in the summation over states. However, it can be reformulated in terms of the expectation value $\langle\beta r^2\rangle$, where $\beta$ is a conventional $4\times4$ Dirac matrix~\cite{Szm02,Kut03}. Evaluated between Dirac bispinors, the operators $r^2$ and $\beta r^2$ have contributions attributed to large and small components of the Dirac bispinors. The inclusion of $\beta$ merely effects a sign change for the small-component contribution, which vanishes in the nonrelativistic limit~\cite{ShiItaBol11}. 

For Yb, we start by considering the differential $M1$ polarizability in the dc limit. Table~\ref{Tab:abbreviatedM1table} presents a breakdown of contributions calculated as detailed in the Supplemental Material~\cite{SMcite}. The final results are compared to the experimental value, which has a $0.1\%$ uncertainty~\cite{mcgrew2018atomic,QZSnote}. As expected, we find that the ${^3P_0}$--${^3P_1}$ ``paramagnetic'' contribution dominates, in part due to a small energy denominator (i.e., the fine structure splitting) in the second-order summation over states. Meanwhile, we find that the diamagnetic contribution amounts to a $\sim\!2\%$ correction, with other contributions being an order of magnitude smaller still. Though sub-dominant, the diamagnetic contribution is non-negligible in the theory-experiment comparison, exemplifying its role in the differential $M1$ polarizability.

\newcommand{\citetab}{\cite{mcgrew2018atomic,QZSnote}}

\begin{table}[]
	\caption{$M1$ differential polarizability, evaluated in the dc limit and at the magic frequency. Theoretical contributions include the ${^3P_0}-{^3P_1}$ paramagnetic (positive energy state) contribution, the diamagnetic (negative energy state) contribution, and other smaller contributions. This is an abbreviated version of a more expansive table presented in the Supplemental Material~\cite{SMcite}, which also includes discussion of theoretical uncertainties. For the dc limit, the final theoretical value is compared to the experimental value. All values are in $10^{-3}$~a.u., where a.u.\ denotes atomic units based on Gaussian electromagnetic expressions.}
	\label{Tab:abbreviatedM1table}
	\begin{ruledtabular}
		\begin{tabular}{lcc}
			contribution		& dc limit			& magic frequency	\\
			\hline
			${^3P_0}-{^3P_1}$	& $5.469$			& $-0.016$			\\
			diamagnetic			& $-0.099$			& $-0.099$			\\
			other				& $0.008$			& $-0.002$			\\
			\hline
			total				& $5.379(10)$		& $-0.116(5)$		\\
			expt.~\citetab		& $5.363(6)$
		\end{tabular}
	\end{ruledtabular}
\end{table}

We next consider the differential $M1$ polarizability evaluated at the magic wavelength (see Table \ref{Tab:abbreviatedM1table} and~\cite{SMcite}). We find that, relative to the dc limit, the ${^3P_0}$--${^3P_1}$ paramagnetic contribution is largely suppressed, a consequence of the lattice photon energy being much greater than the fine structure splitting. Meanwhile, the dc value for the diamagnetic contribution can be directly applied for the magic wavelength case, as the photon energy is significantly below the energy associated with electron-positron pair production. It follows that the diamagnetic contribution becomes the dominant contribution for the differential $M1$ polarizability at the magic wavelength. Further, evaluating and including the differential $E2$ polarizability at the magic wavelength~\cite{SMcite}, we obtain the theoretical result $\tilde{\alpha}_{M1E2}=-1.9(5)\times10^{-18}$, in good agreement with the experimental results (Table~\ref{table_summary}). Finally, using formalism described in Ref.~\cite{PorSafKoz18} we found $\tilde{\beta}~=-2.3 \times 10^{-21}$ in the CI+all-order approximation. In two dominant terms, we replaced the theoretical denominators with more correct experimental ones, that strongly affect the result.  We consider the result an order of magnitude estimate.

\textit{Summary.}--With multi-ensemble operation and imaging, we realize a complete lattice light shift evaluation of a standard retro-reflected 1D OLC using modest trap depths. Our independent evaluation provides valuable atomic coefficients for Yb OLCs while also demonstrating novel techniques for the evaluation of both $\frac{\partial \tilde{\alpha}_{E1}}{\partial \nu}$, $\tilde{\beta}$, and $\tilde{\alpha}_{M1E2}$~\cite{ovsiannikov2013multipole}. Finally, the experimental and theoretical results from this Letter further validate the recent consensus on the origin of the disagreement on the sign and magnitude of the multipolarizability term $\tilde{\alpha}_{M1E2}$. 

\textit{Acknowledgments.}--We thank K. Kim and A. Staron for careful reading of the manuscript. T.B. acknowledges insightful conversations with A. Goban and R. Hutson. The experimental work was supported by NIST, ONR, and NSF QLCI Grant No. 2016244 and Grant No. 2012117 (KG). T.B. acknowledges support from the NRC RAP. The theoretical work has been supported in part by the US NSF Grants  No. PHY-2309254,  OMA-2016244, US Office of Naval Research Grant No. N00014-20-1-2513, and by the European Research Council (ERC) under the Horizon 2020 Research and Innovation Program of the European Union (Grant Agreement No. 856415). Calculations in this work were done through the use of Information Technologies resources at the University of Delaware, specifically the high-performance Caviness and DARWIN computer clusters.

\bibliography{references.bib}

\newpage
\section{Appendix A: Born-Oppenheimer + WKB Approximation}

The lattice light shift model in the main text follows a standard harmonic basis treatment~\cite{ushijima2018operational}. While it gives important physical intuition, these models are known to break down at higher temperatures as they fail to capture axial-radial couplings~\cite{beloy2020modeling}. Considering our radial temperature of $\sim 600 $ nK ($\sim1~\mu$K in the running wave measurements), we elect to perform an additional analysis using a Born-Oppenheimer+WKB treatment (BO+WKB) which better captures axial-radial couplings~\cite{beloy2020modeling}. In this treatment the lattice light shift is given by

\begin{equation}
	\begin{aligned}
		\frac{\delta \nu_{LS}(u,\delta_L,n_z,T_r)}{\nu_{c}} \approx& -\sum_{n_z} W_{n_z}\bigg[ \frac{\partial \tilde{\alpha}_{E1}}{\partial \nu}\delta_L  X(n_z,u_0,T_r) u_0\\
		& +\tilde{\alpha}_{M1E2} Y(n_z,u_0,T_r) u_0 \\
		& + \tilde{\beta} Z(n_z,u_0,T_r) u_0^2 \bigg ],
	\end{aligned}
	\label{equation_BO+WKB}
\end{equation}

\noindent
where $W_{n_z}$ is an $n_z$ band weight and $u_0$ is the peak trap depth normalized by $E_R$. $X(n_z,u_0,T_r)$, $Y(n_z,u_0,T_r)$, and $Z(n_z,u_0,T_r)$ are trap depth reduction factors which are numerically calculated~\cite{beloy2020modeling}. As presented in Table~\ref{table_BOWKB}, we find good agreement between models, but note a 1-$\sigma$ discrepancy of $\frac{\partial \tilde{\alpha}_s{E1}}{\partial \nu}$. We note that future evaluations with improved uncertainties will likely need to utilize colder temperatures to continue to employ the harmonic basis model.

\begin{table}[]
	\begin{center}
		\caption{Comparison of experimental results as derived from either the harmonic (Eq.~(\ref{equation_lattice_light_shift})) or BO+WKB (Eq.~(\ref{equation_BO+WKB})) treatment. }
		\label{table_BOWKB}
		\begin{tabular}{c c c} 
			\hline \hline  
			Coefficient & Harmonic Basis    & BO+WKB\\ [.5ex] 
			\hline  
			\rule{0pt}{1.\normalbaselineskip} $\frac{\partial \tilde{\alpha}_{E1}}{\partial \nu}$ ($10^{-20}$/MHz) & 4.21(10)& 4.31(9) \\ 
			
			$\tilde{\beta}$ ($10^{-21}$) & $-1.7(4)$& $-2.0(6)$  \\
			
			$\tilde{\alpha}_{M1E2}^{\text{Experiment}}$ $(10^{-18})$ & $-1.41(9)$& $-1.45(8)$\\

			$\nu_{E1}$ (MHz) & $394~798~266.9 (26)$ &  $394~798~266.3 (30)$ \\
			
			$\nu'_r(0)$ (MHz) & $394~798~300.4 (18)$ &  $394~798~300.0 (25)$ \\ 
			\hline \hline 
		\end{tabular}
	\end{center}
\end{table}

\end{document}


\newcommand{\etal}{{\it et al.}}
\newcommand{\pprime}{{\prime\prime}}
\newcommand{\ppprime}{{\prime\prime\prime}}
\newcommand{\eref}[1]{Eq.~(\ref{#1})}
\newcommand{\tref}[1]{Table~\ref{#1}}



\preprint{}
\title{Supplementary Material for Lattice Light Shift Evaluations In a Dual-Ensemble Yb Optical Lattice Clock}

\author{Tobias Bothwell\textsuperscript{1,*}, Benjamin D. Hunt\textsuperscript{1,2,*}, Jacob L. Siegel\textsuperscript{1,2,*}, Youssef S. Hassan\textsuperscript{1,2},
	Tanner Grogan\textsuperscript{1,2}, Takumi Kobayashi\textsuperscript{1,4}, Kurt Gibble\textsuperscript{1,5}, Sergey G. Porsev\textsuperscript{6}, Marianna S. Safronova\textsuperscript{6}, Roger C. Brown\textsuperscript{1}, Kyle Beloy\textsuperscript{1}, Andrew D. Ludlow\textsuperscript{1,2,3,$\dagger$}}

\affiliation{\textsuperscript{1}National Institute of Standards and Technology, 325 Broadway, Boulder, Colorado 80305, USA}
\affiliation{\textsuperscript{2}University of Colorado, Department of Physics, Boulder, Colorado 80309, USA}
\affiliation{\textsuperscript{3}Electrical, Computer \& Energy Engineering, University of Colorado, Boulder, Colorado, USA}
\affiliation{\textsuperscript{4}National Metrology Institute of Japan (NMIJ), National Institute of Advanced Industrial Science and Technology (AIST), 1-1-1 Umezono, Tsukuba, Ibaraki 305-8563, Japan}
\affiliation{\textsuperscript{5}Department of Physics, The Pennsylvania State University, University Park, Pennsylvania 16802, USA}
\affiliation{\textsuperscript{6}Department of Physics and Astronomy, University of Delaware, Newark, Delaware 19716, USA}
\def\thefootnote{*}\footnotetext{These authors contributed equally to this Letter.}\def\thefootnote{\arabic{footnote}}
\def\thefootnote{$\dagger$}\footnotetext{Contact author: andrew.ludlow@nist.gov}\def\thefootnote{\arabic{footnote}}
\date{\today}
\maketitle


\newcommand{\e}[1]{\ensuremath{\times10^{#1}}}
\newcommand{\SP}{\multirow{3}{*}{\makebox[2.5mm]{P}
$\left\{\begin{array}{l}
\text{valence}\\
\text{core}\\
\text{valence-core}
\end{array}\right.$
}}
\newcommand{\PP}{\multirow{5}{*}{\makebox[2.5mm]{P}
		$\left\{\begin{array}{l}
			\text{valence},~^3\!P_0 -\,^3\!P_1\\
			\text{Schwinger QED}\\
			\text{valence, others}\\
			\text{core}\\
			\text{valence-core}
		\end{array}\right.$
}}
\newcommand{\D}{\multirow{3}{*}{\makebox[2.5mm]{D}
		$\left\{\begin{array}{l}
			\text{valence}\\
			\text{core}\\
			\text{RPA correction}
		\end{array}\right.$
}}

\section{Preparation of Motional State Ensembles}

\begin{figure*}
	\centering
	\includegraphics[width=0.98\textwidth]{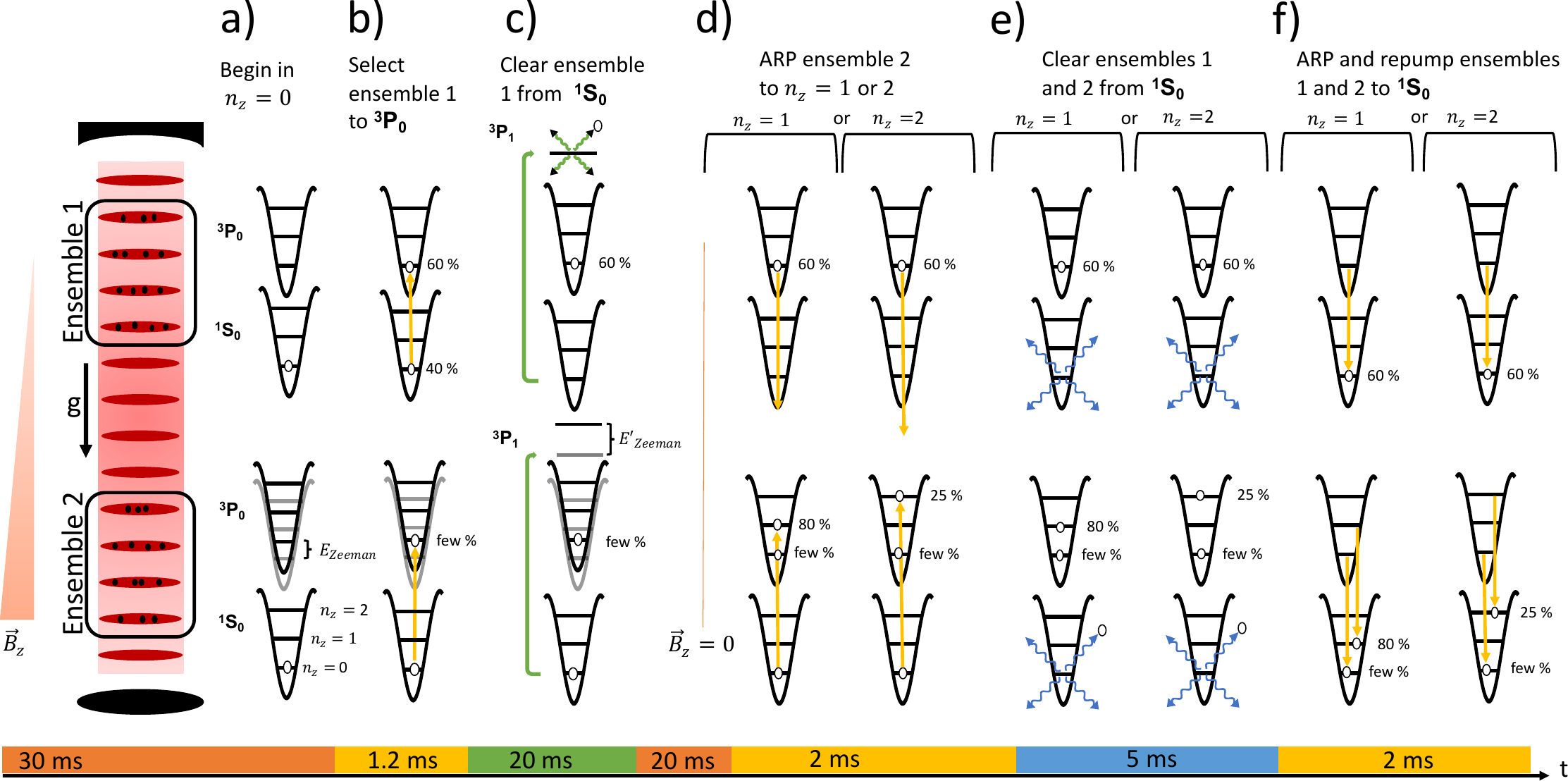}
	\caption{The process used to prepare ensemble 1 in $n_z\approx0$ and ensemble 2 in $n_z\approx 1$ or $2$. Each step outlined in the text is shown sequentially as \textbf{a)} through \textbf{f)}, where the duration of each step is shown at the bottom. The percentages denote the fraction of the original sample, yellow arrows denote 578-nm laser pulses, green 556-nm laser pulses, and blue 399-nm scattered photons. $E_{Zeeman}$ is the Zeeman shift of the clock transition due to the vertical magnetic field at the position of ensemble 2, shown here for simplicity affecting \textsuperscript{3}P\textsubscript{0} only, and $E'_{Zeeman}$ is the Zeeman shift of the 556-nm transition, again shown only for \textsuperscript{3}P\textsubscript{1} for simplicity. The levels unperturbed by the Zeeman shift are shown with lighter shading. For steps \textbf{d) }through \textbf{f)} either the $n_z=1$ or the $n_z=2$ excitation option is performed, depending on the desired final motional state. These final motional state options only differ by performing the ARP pulse in \textbf{d)} on the $\Delta n_z=1$ or $\Delta n_z=2$ sideband. }
	\label{fig:select}
\end{figure*}

We use the vertical magnetic field gradient of the MOT coils to spectroscopically resolve ensembles 1 and 2, allowing preparation of ensemble 1 in $n_z\approx0$ and ensemble 2 in $n_z\approx1$ or $2$. We begin by preparing both ensembles in $n_z \approx 0$. The remaining preparation process, outlined below, is shown in Figure \ref{fig:select} for steps \textbf{a)} through \textbf{f)}.
For \textbf{a)}, we apply a magnetic field gradient such that B\textsubscript{vertical} is $B_z \approx 0$ G for ensemble 1, and $B_z \approx 5$ G for ensemble 2. 
\textbf{b)} A 1.2 ms $\pi$-pulse prepares $\approx60\%$ of ensemble 1 in the excited clock state, which is nominally unperturbed by the Zeeman shift, while selecting only a few percent of ensemble 2. 
A 1.2 ms pulse is experimentally chosen to optimize the pulse fidelity for ensemble 1 while minimizing the excitation of ensemble 2, which is off-resonant due to its large Zeeman shift.
Unintentional excitation of ensemble 2 results in imperfections in $n_z$ preparation, decreasing the final $<n_z>$. With the vertical magnetic field gradient still on, in \textbf{c)} we apply a 20-ms resonant 556-nm pulse to clear the $\approx40\%$ of atoms remaining in the ground clock state of ensemble 1, with ensemble 2 again shielded by the difference in applied magnetic field. 
Ground state atoms in ensemble 1, with approximately zero magnetic field, are resonant with the 556-nm transition and are heated out of the lattice.  We typically see this clearing pulse is $>97\%$ effective for ensemble 1, while retaining $\approx95\%$ of ensemble 2. The magnetic field gradient is then turned off at the beginning of \textbf{d)} and given 20 ms to settle before an adiabatic rapid passage (ARP) pulse on the first (or second)-order blue sideband is applied to both ensembles. 
The excited atomic sample in ensemble 1 is in $n_z=0$, a dark state for the frequency of the applied ARP pulse. Ensemble 2 is promoted to the exited clock state and $n_z=1$ ($n_z=2$) depending on excitation on the first (second) blue sideband, with $\sim$80\% (25\%) transfer efficiency. 

In \textbf{e)}, any remaining ground state atoms in either ensemble are removed via a 5 ms 399-nm pulse.
Finally for \textbf{f),} ensemble 1 and ensemble 2, both in the excited clock state in $n_z=0$ and $n_z=1$ or $2$ respectively, are excited to the ground state via an ARP pulse on the carrier transition followed by 1388-nm light to repump any atoms not moved by the carrier ARP pulse. 
Optical pumping is performed before proceeding to differential spectroscopy of the ensembles.

With our current control capabilities, a non-negligible fraction ($\sim 25 \%$) of the final sample of ensemble 2 is found to be in motional states other than desired. 
This is dominated by the accidental selection of a few percent of $n_z=0$ atoms to \textsuperscript{3}P\textsubscript{0} in  \textbf{b)}.
We note that separating the samples by $\sim1$-mm, a distance that is several times farther than that used in this Letter, reduced this unintended selection to $<$1\%, suggesting it is due to off-resonant excitation. A less-significant source of motional state impurity is optical pumping after \textbf{f)}, where scattered photons may change the motional state. The selection and spin polarization process lead to some heating of ensemble 1 to $n_z\sim0.13$. We also search for $n_z=3(2)$ population in the $n_z=2(1)$ sample in ensemble 2, via sideband spectroscopy. No excitation was found for the $\Delta n_z=-3(2)$ sideband, so we do not assign any population to $n_z>2(1)$.

\section{Fitting Sideband Spectra to Extract $<n_z>$}
Because the method to prepare one ensemble in $n_z=0$ and the second in $n_z=1,2$ has imperfect fidelity, we use sideband spectroscopy to extract the relative populations of $n_z = 0,1,$ or 2 of each ensemble. With Sisyphus cooling yielding sufficiently low radial temperatures, sideband excitations of individual lattice bands are resolved \cite{PhysRevLett.129.113202}, revealing that the standard harmonic model treatment \cite{blatt2009sideband} does not capture the energies of highly excited lattice bands. To address these concerns we develop a treatment based on \cite{beloy2020modeling}. To begin, we  first find the peak optical trap depth ($U_{max}$) from the corner frequency of the harmonic model in the usual way. To more correctly capture higher-order band spacings, we next evaluate each band's corner frequency, $U_{n_z}(\rho)$, using Mathieu functions, as 

\begin{equation}
	\begin{aligned}
		U_{n_z}(\rho) =& E_R \bigg [  b_{n_z+1}\bigg( \frac{D(\rho)}{4}\bigg )  - \frac{D(\rho)}{2}\bigg ] ,  
	\end{aligned}
	\label{equation_mathieu_sidebands}
\end{equation}

\noindent
where $b_r(q)$ is the characteristic value for the odd Mathieu function \cite{beloy2020modeling}. Here, $D(\rho) = (U_{max}/E_{R}) e^{-\rho^2/2\omega_0^2}$  gives the radial potential for radius $\rho$ and waist $\omega_0$. The differences of $n_z$ dependent solutions of equation (\ref{equation_mathieu_sidebands}) then give the corner frequency for each blue-sideband (BSB) corner frequency $\nu_{BSB}$, i.e. $\nu_{BSB}(n_z=1 \rightarrow 2) = (U_{2}(0)-U_{1}(0))/h$ where $h$ is Planck's constant.

We now consider thermally sampling $\rho$. For each $n_z$ value simulated, we construct an evenly sampled set of $\sim$100 frequencies from the corner frequency $\nu_{BSB}(n_z \rightarrow n_z+1)$ to 20 kHz below the corner frequency, sufficient to capture experimentally observed band excitation. A single, global radial temperature for all bands is then used to provide Boltzman weighting ($W(n_z,\rho)$) for each band's sampled frequencies, taking their energies to be $E(\rho) = U_{n_z}(\rho)-U_{n_z}(0) $. Atoms with larger $\rho$ values (higher radial temperatures) experience lower effective trap depths. To capture this, we associate a Lamb-Dicke parameter $\eta^2(n_z,\rho) = E_R/U_{n_z}(\rho)$ with each $U_{n_z}(\rho)$ value. This in turn allows us to associate each sampled $n_z$ and $\rho$ value with an effective Rabi frequency for the $n_z \rightarrow n_z+1$ BSB transition as

\begin{equation}
	\begin{aligned}
		\Omega_{BSB}(n_z,\rho) = \Omega_0 e^{-\eta^2(n_z,\rho)/2} \eta(n_z,\rho) \sqrt{\frac{n_z !}{(n_z+1)!}} L^1_{n_z} \bigg ( \eta^2 (n_z, \rho ) \bigg ),
	\end{aligned}
	\label{equation_rabi_LD}
\end{equation}

\noindent
where $\Omega_0$ is the bare Rabi frequency and $L^\alpha_{n}(X)$ is the generalized Laguerre polynomial \cite{leibfried2003quantum}. 

For a laser frequency $\nu_i$, the excitation fraction ($P_e$) is found via 

\begin{equation}
	\begin{aligned}
		P_e(\nu_i,U_{max},\Omega_0,T_r,A_0,A_1...B) = \frac{1}{2}\sum_{n_z} A_{n_z} \sum_k \frac{W_{n_z,k}\Omega_k^2}{\Omega_k^2+(\nu_{BSB}^{n_z,k}-\nu_i)^2} + B.
\end{aligned}
\label{equation_sim_sideband}
\end{equation}

\begin{figure}
	\includegraphics[width=0.49\textwidth]{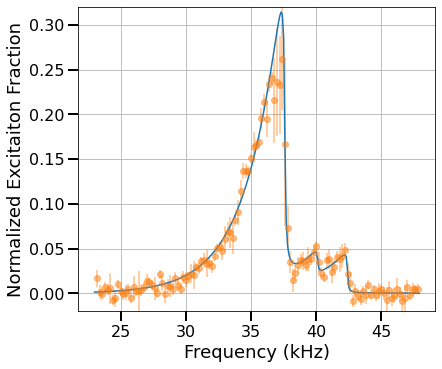}
	\caption{A representative sideband fit (blue line) for ensemble 2 prepared ideally in $n_z=2$ at 123 $E_R$ is shown. The error bars of the data points are the standard deviation of two points for each frequency added in quadrature with a readout noise of 0.5\%. We see an $n_z=0$ population of 19.2(26)\%, $n_z=1$ of 7.7(11)\% and $n_z=2$ of 73.1(27)\% Not shown is the sideband scan for ensemble 1, ideally prepared in $n_z = 0$, which yields $U_{max}$ for the fit and has a residual population in $n_z=1$ of $8(2)\%$. }
	\label{figure_sideband_scan_with_fit}
\end{figure}

\noindent
We Fit Eq.~(\ref{equation_sim_sideband}) to the sideband spectra, determining the radial temperature $T_r$, individual $n_i$ population weights $A_i$, Rabi frequency $\Omega_0$, and excitation offset $B$ given laser frequency $\nu_i$ and max trap depth $U_{max}$ as inputs. The index $k$ accounts for the thermal sampling of each $n_z$ specific set of corner frequencies, $\nu_{BSB}^{n_z,k}$, having a Boltzmann weight $W_{n_z,k}$. We implicitly assume time-averaged dynamics and single particle physics free of dephasing. We find this model reproduces the observed sideband spectra as shown in Fig. \ref{figure_sideband_scan_with_fit}. We assign error from both least-squares fitting uncertainties and disagreement in $<n_z>$ from a less-sophisticated fitting of the amplitudes of the data at the corner frequencies, typically corresponding to an $\delta <n_z> \approx 0.1$.

\section{Model-Dependent Reduction Factors}

In the main text we report atomic coefficients utilizing a harmonic basis model \cite{ushijima2018operational}. As discussed in both Appendix A and Ref.~\cite{beloy2020modeling}, accounting for radial temperatures can introduce model dependent biases when atoms are not sufficiently cold relative to the trap depth. We thus verify in Appendix A that at the current level of uncertainty our results are consistent with a more thorough treatment (BO+WKB) which respects radial-axial couplings. For additional reference, we give example temperatures and trap depth reduction factors for both models in Table~\ref{Tab:reductions}.

\begin{table}[h]
	\caption{Exemplification of light shift model differences, comparing the ``harmonic basis'' model (Ref.~\cite{ushijima2018operational,kim2023evaluation} and Equation (1) of the main text) to the BO+WKB model~\cite{beloy2020modeling}. Expressions for the factors $X(n_z,u_0,T_r)$, $Y(n_z,u_0,T_r)$, and $Z(n_z,u_0,T_r)$ for the respective models can be found in Ref.~\cite{beloy2020modeling}. The four lines tabulated here correspond to the four data points appearing in Fig.~2 of the main text. We note that the abscissa in Fig.~2 of the main text is $u^\prime=u_0W_{n_z=0}X(n_z,u_0,T_r)$, computed with the harmonic basis model. In Ref.~\cite{beloy2020modeling}, the harmonic basis model is referred to as the modified Ushijima~{\it et al.}\ model.}
	\label{Tab:reductions}
	\begin{ruledtabular}
		\begin{tabular}{cccccccc}
			& 
			& \multicolumn{2}{c}{$X(n_z=0,u_0,T_r)$}
			& \multicolumn{2}{c}{$Y(n_z=0,u_0,T_r)$}
			& \multicolumn{2}{c}{$Z(n_z=0,u_0,T_r)$}
			\\
			\cline{3-4}\cline{5-6}\cline{7-8}
			\vspace{-3mm}\\
			$u_0$	& $T_r$~(nK)	
			& harmonic basis	& BO+WKB	
			& harmonic basis	& BO+WKB	
			& harmonic basis	& BO+WKB	\\
			\hline\vspace{-3mm}\\
			$56.8$	& $650$		& $0.832$	& $0.785$	& $0.0627$	& $0.0608$	& $0.708$	& $0.645$	\\
			$66.4$	& $550$		& $0.863$	& $0.838$	& $0.0589$	& $0.0580$	& $0.756$	& $0.719$	\\
			$86.2$	& $600$		& $0.881$	& $0.864$	& $0.0520$	& $0.0515$	& $0.786$	& $0.759$	\\
			$112.2$	& $720$		& $0.892$	& $0.879$	& $0.0457$	& $0.0454$	& $0.804$	& $0.781$		
		\end{tabular}
	\end{ruledtabular}
\end{table}

\section{Cold Collisions}

The cold collision shift, arising from atomic interactions, can contaminate lattice light shift measurements. A major benefit of evaluating $\frac{\delta \tilde{\alpha}_{E1}}{\delta \nu}$ via interleaved lattice frequencies is identical atomic preparation and therefore cold collision shifts. For differential measurements between atom samples, identical atom number preparation and density are not feasible.

Ratchet loading allows the relative density between two samples to be varied. By controlling the efficiency of the shelving pulse, it is possible to obtain large differences in atom number between the two ensembles. After longitudinal state preparation, synchronous spectroscopy of these two samples provides a direct measurement of the differential frequency shift as a function of the atom number difference between samples. While this method provides rapid evaluation compared to interleaved comparisons, it can only provide a correction at the 10\% level of accuracy due to a small difference (10\%) in the sample lengths between ensemble 1 and 2.

For the Sisyphus cooled samples used in the differential motional state study (Figure 4 of the main text), density shifts were evaluated by measuring the frequency difference between otherwise identically prepared ensembles. Critically, when both ensembles are prepared in the same $n_z$ state, the preparation fidelity is far higher than for the differential $n_z$ preparation. For $n_z =$ 0, 1, and 2 the cold collisional shift in fractional frequency units of $10^{-18}/1000$ atoms was measured to be -6.14 (0.42), -5.52(0.63) and -4.11(0.42) at trap depths of $\sim$124, 129, and 135 $E_{R}$ respectively. To apply corrections to the data of Figure 4, each frequency difference had a $<n_z>$ weighted cold collision shift correction based on the fitted sideband spectra. Shifts were scaled to the correct depth using a $U^{5/4}$ scaling \cite{nicholson2015new,nicholson2015systematic}, due to challenges associated with measuring vanishingly small cold collisional shifts at shallow trap depths. For thermal samples without sideband cooling, a $U^{3/2}$ scaling may hold. Additional uncertainty was included to  conservatively account for the difference between these models. The data in Figure 4 is thus well described at large trap depths where cold collision shifts are dominant and the extrapolation of the coefficients is small relative to coefficient, and at shallow depths where the applied corrections are vanishingly small. For context, the largest applied cold collision correction in this data is 3.5(8)$\times 10^{-18}$. 

For the auxiliary running wave measurements of Fig. 3, the atoms were not Sisyphus cooled as discussed in the text. In this regime the cold collisional shift has been previously evaluated \cite{mcgrew2018atomic}. Additionally, care was taken at the largest depths to ensure $<$200 atom number differences between regions with $\approx 1000$ atoms. Under such conditions we expect that the magnitude of  cold collisonal correction to be of order $1\times10^{-19}$, providing an uncertainty on the running wave magic frequency an order of magnitude smaller than the uncertainty of the result quoted in the text. Similarly, we estimate a difference of trapping volumes between regions due to the running wave to be at the 1\% level, which can be neglected here.

\section{Multi-ensemble Frequency Comparison}

Extracting the frequency difference between multiple ensembles in a single apparatus from the measured excitation fractions of each ensemble requires converting excitation fractions to frequencies. To do so, we utilize a Rabi line-shape  \cite{steck2007quantum}

\begin{equation}
	\label{eqn:Rabi}
	P_i=C_i \frac{\Omega^2}{\Omega^2+\delta_i^2}\mathrm{sin}^2\left(\frac{T_\pi}{2}\sqrt{\Omega^2+\delta_i^2}\right),
\end{equation}

\noindent
where $P_i$ is the excitation fraction of ensemble $i$, $C_i$ the contrast, $\Omega$ is the Rabi frequency, $T_\pi$ the pulse time optimized for a $\pi$-pulse, and $\delta_i$ the detuning from resonance. For clock operation the Rabi lineshape is probed on both sides of the lineshape with the frequency difference given by the full-width-at-half-max. Each ensemble's contrast is then taken as twice the average excitation from an experimental run, capturing in-situ differences in contrast. 

As in standard OLC operation \cite{ludlow2008strontium}, we probe the clock transition of opposite magnetic states ($m_F = \pm 1/2$) by interrogating both sides of each spin state's Rabi lineshape. For a single ensemble and single state, we can then take the difference in excitation fraction as measured on opposite sides of the lineshape and, using Eq.~(\ref{eqn:Rabi}), map the difference to a frequency offset from the true lineshape center. The difference in this frequency offset between ensembles then provides a frequency difference between ensembles with common-mode laser noise rejected, enhancing the measurement stability.  

To bound potential errors from the conversion of excitation fractions to frequency, we perform simulations of our analysis using data generated from the analytical line-shape of Eq.~(\ref{eqn:Rabi}), including laser noise and quantum projection noise \cite{ludlow2015optical}. We find that this method generates errors of the frequency difference between the ensembles linear in the error in the contrast. For example, a $2\times10^{-17}$ shift measured between $n_z=0$ and $n_z=2$ would incur a $1\times10^{-18}$ error for a 5\% error of the contrast. Evaluation of the contrast from clock locks, line scans, and lineshape fitting bounds the contrast uncertainty to $2\%$. Even with a perfect reproduction of contrast, the simulations suggest a potential biasing of results (similar to a servo-error in standard clock evaluations) at the level of $<1\%$ of the measured shift. To account for these effects, an additional 2\% uncertainty relative to the measured differential frequency shifts is added in quadrature with statistical uncertainties for all measurements presented in the main text.

\section{Running Wave Light Shift}

The analysis of the running wave light shift included two additional effects not discussed in the main text. First, the running wave light shift neglected a higher-order term $\delta \nu_{R}(u_r)/\nu_c \approx-\tilde{\beta}u_r^2$. For the running wave trap depths of $u_r\sim 10$ $E_R$ this resulted in a 0.4(1) MHz correction to $\nu_r(0)$, which is included in the reported value. Second, the ellipticity of the lattice was bounded to be $\le10\%$, which we take as an extra uncertainty for $\tilde{\beta}$, added in quadrature with the statistical uncertainty. 

\section{Theoretical Calculations}

\subsection{General formalism}
The ac $2^K$-pole polarizability of the $|0\rangle$ state can be expressed (in atomic units $\hbar=m=|e|=1,\,c \approx 137$)
as~\cite{PorDerFor04}
\begin{eqnarray}
	\alpha_{\lambda K}(\omega) &=& \frac{K+1}{K\,[(2K-1)!!]^2} \left(\frac{\omega}{c}\right)^{2K-2}  
	 \sum_n \frac{\Delta E_n | \langle n|(T_{\lambda K})_0 |0 \rangle |^2}{(\Delta E_n)^2-\omega^2} .
	\label{Qlk}
\end{eqnarray}
Here, $\Delta E_n \equiv E_n - E_0$, $\lambda$ distinguishes between electric, $\lambda = E$, and magnetic, $\lambda =M$, multipoles,
and $(T_{\lambda K})_0$ is the $0$ component of the operator $T_{\lambda K}$ in spherical coordinates, where $T_{E1} \equiv D$, $T_{M1} \equiv \mu$, and $T_{E2} \equiv Q_2$.
These many-electron operators are expressed as the sum of the single-electron operators. For example,
${\bm \mu} = \sum_{i=1}^N {\bm \mu}_i$, where $N$ is the number of electrons in the atom. The sum over $n$ in \eref{Qlk} includes the positive- and negative-energy states, labeled in the following by $n^+$ and $n^-$, respectively.

As discussed in Ref.~\cite{PorKozSaf23}, when calculating the $E2$ polarizabilities,
\begin{eqnarray}
	\alpha_{E2}(\omega) &=& \frac{1}{6} \left(\frac{\omega}{c}\right)^{2}
	\sum_n \frac{\Delta E_n | \langle n| Q_{20} |0 \rangle |^2}{(\Delta E_n)^2-\omega^2} ,
	\label{E2}
\end{eqnarray}
the contribution of intermediate negative-energy states is negligible.

For ac $M1$ polarizabilities, the negative-energy states instead give the dominant contribution to
the $M1$ polarizabilities of both clock states at the magic frequency. To calculate $M1$ polarizabilities, we use the expression
derived in Ref.~\cite{PorKozSaf23}. Neglecting terms $\sim 1/c^4$, we have
\begin{eqnarray}
	\alpha_{M1}(\omega) &\approx&
	2 \sum_{n^+} \frac{\Delta E_{n^+}}{(\Delta E_{n^+})^2-\omega^2} |\langle n^+ |\mu_0| 0 \rangle|^2 
	- \frac{1}{6 c^2}\, \langle 0 | r^2  |0 \rangle ,
	\label{aM1_2}
\end{eqnarray}
where the first term in \eref{aM1_2} is associated with the contribution of positive-energy states and the second term is associated
with the contribution of negative-energy states.
\subsection{Method of calculation}
We consider Yb as an atom with two valence electrons above a closed shell core and perform calculations within the framework of methods that combine configuration interaction (CI) with (i) many-body perturbation theory (MBPT)~\cite{DzuFlaKoz96} and (ii) the linearized coupled-cluster method~\cite{SafKozJoh09}. In these methods, the energies and wave functions are found from the multiparticle
Schr\"odinger equation
\begin{equation}
	H_{\mathrm{eff}}(E_n) \Phi_n = E_n \Phi_n,
	\label{Heff}
\end{equation}
where the effective Hamiltonian is defined as
\begin{equation}
	H_{\mathrm{eff}}(E) = H_{\mathrm{FC}} + \Sigma(E).
	\label{Heff1}
\end{equation}
Here, $H_{\mathrm{FC}}$ is the Hamiltonian in the frozen core (Dirac-Hartree-Fock) approximation, and $\Sigma$ is the energy-dependent correction, which takes into account virtual core excitations in the second order of the perturbation theory (the CI+MBPT method), or to all orders (the CI+all-order method).
\subsection{Calculation of $\alpha_{M1}$ and $\alpha_{E2}$}
To calculate the first (paramagnetic) term in \eref{aM1_2}, we used the sum-over-states approach, including the contribution of several
low-lying intermediate states. We also found the core part and the small valence-core contribution, restoring the Pauli principle.
The results obtained in the framework of the CI+all-order approximation and labeled by ``P'' (``valence'', ``core'', and ``valence-core'') are presented in \tref{Tab:breakdown}. The paramagnetic part
of the $^3\!P_0$ $M1$ polarizability is mostly determined by the contribution of the intermediate $6s6p\,^3\!P_1$ state.
We separated this contribution in the row labeled ``valence, $^3\!P_0 -\, ^3\!P_1$.'' The contribution of other valence
intermediate states is given in the row ``valence, other.'' The quantum-electrodynamical (QED) correction to the
$\langle ^3\!P_0 ||\mu|| ^3\!P_1 \rangle$ matrix element (ME), discussed below in more detail, is presented in the row labeled
``Schwinger QED.''

The calculation of the diamagnetic contribution to the $M1$ polarizability is reduced to determining a matrix element
$\langle 0 | r^2  |0 \rangle$, where  $|0 \rangle$ is either the $^1\!S_0$ or $^3\!P_0$ state. This ME can be divided into its
valence and core parts as $$\langle 0 | r^2 |0 \rangle = \langle 0 |r^2|0 \rangle_v + \langle 0 | r^2 |0 \rangle_c.$$
To find the valence parts of $\langle ^1\!S_0 | r^2 | ^1\!S_0 \rangle$ and $\langle ^3\!P_0^o | r^2 | ^3\!P_0^o \rangle$ and estimate their uncertainties, we carried out the calculation using the CI+MBPT and CI+all-order methods.
The core part was found in the single-electron approximation,
\begin{eqnarray}
	\langle 0 | r^2  |0 \rangle_c = \sum_{a=1}^{N_c} \langle a | r_a^2 | a \rangle ,
	\label{r2core}
\end{eqnarray}
where $|a\rangle$ is the single-electron wave function of the $a$th core electron and $N_c$ is the number of core electrons.
The CI+all-order results are given in \tref{Tab:breakdown} in the panel labeled ``D.''

The correlation corrections to the expectation values of the operator $r^2$ arise from the correlation corrections to the wave functions and the corrections to the operator. The latter include the random-phase approximation (RPA) and smaller corrections
beyond the RPA. The RPA correction (given in the row ``RPA correction'') is less than 1.5\% for both MEs.
The smaller corrections  were not calculated, but, as we observed for Sr~\cite{PorKozSaf23}, the sum of these corrections
tend to cancel the RPA correction, giving in total a small contribution.

To find $\alpha_{E2}$, we used the Sternheimer~\cite{Ste50} or Dalgarno-Lewis~\cite{DalLew55} method and solve the
inhomogeneous equation
\begin{eqnarray}
	(H_{\rm eff} - E_0 \pm \omega)\, |\delta \phi_{\pm} \rangle = Q_{20}\, |0 \rangle,
	\label{inhom}
\end{eqnarray}
where $H_{\rm eff}$ is the effective Hamiltonian determined by \eref{Heff1}. Then, \eref{E2} can be written as
\begin{eqnarray}
	\alpha_{E2}(\omega) &=& \frac{1}{12} \left(\frac{\omega}{c}\right)^{2}
	\left[ \langle 0| Q_{20} |\delta \phi_{+} \rangle + \langle 0| Q_{20} |\delta \phi_{-} \rangle \right] .
	\label{E2_2}
\end{eqnarray}
The results obtained in the CI+all-order approximation are displayed in \tref{Tab:E2}.

\tref{Tab:combined} summarizes the final results for the $M1$ and $E2$ polarizabilities of the clock
states. To estimate uncertainties, we presented the results obtained in the CI+MBPT and CI+all-order approximations.
The uncertainty in $M1$ polarizabilities is well controlled as it mostly comes from $\langle 0 |r^2| 0 \rangle$ MEs.
As this is a single ME, the uncertainty can be estimated in a standard way as the difference of the CI+all-order and CI+MBPT
values. In the case of the dc M1 differential polarizability, we also take into account the uncertainties of the paramagnetic contributions listed in Table~\ref{Tab:breakdown}. Total uncertainty in the dc value is taken to be the sum of estimated uncertainties in all contributions which are as follows: valence $^3$P$_0$-$^3$P$_1$  $5.444(2)\times10^{-3}$; Schwinger QED $2.5(2)\times10^{-5}$; valence (other) $7(1)\times10^{-6}$; valence (diamagnetic) $-9.9(5)\times10^{-5}$.

The uncertainties of the $E2$ polarizabilities cannot be estimated in this way because, according to our calculation, the
high-lying discrete states and continuum contribute about 50\% to the total value of the $^3\!P_0$ $E2$ polarizability.
This suggests that a large number of intermediate states with the open 4$f$-shell can contribute. We do not reproduce such states
in the framework of our method and have no benchmark to assess such a case. As a result, we conservatively estimate the uncertainty
of this $E2$ polarizability to be 20-30\%. For the $^1\!S_0$ state, the contribution of the high-lying states is about 15\% and we estimate the uncertainty of the $^1\!S_0$ $E2$ polarizability to be 10-15\%.

\begin{table}
\caption{Computed $M1$ polarizabilities in the CI+all-order approximation for the Yb clock states in the dc limit and at the magic frequency. Contributions are partitioned into paramagnetic and diamagnetic contributions, labeled by P and D. These correspond to ``positive energy state'' and ``negative energy state'' contributions, respectively, in the language of
Refs.~\cite{WuShiNi23,PorKozSaf23}. For the $6s6p\,^3\!P_0$ state, the
$^3\!P_0 -\, ^3\!P_1$ fine structure contribution is listed separately from other ``valence'' paramagnetic contributions to highlight its relative importance, especially in the dc limit. For this contribution, the theoretical $M1$ matrix element is taken together with the experimental fine structure splitting for $^{171}$Yb~\cite{JonKanMcF23}. The Schwinger QED (anomalous electron magnetic moment) correction is accounted for separately (see text). For the dc case, the final differential result is compared to the experimental value for $^{171}$Yb (see text). All values are in atomic units based on Gaussian electromagnetic expressions.}
	\label{Tab:breakdown}
	\begin{ruledtabular}
		\begin{tabular}{llcc}
			&			& dc limit			& magic frequency	\\
			\hline\vspace{-3mm}\\
			$6s^2\,\,^1\!S_0$
			& \SP		& $1.5\e{-8}$		& $1.6\e{-8}$		\\
			&			& $1.9\e{-7}$		& $1.9\e{-7}$		\\
			&			& $-3\e{-12}$		& $-3\e{-12}$		\\
			\vspace{-3mm}\\
			& \D		& $-3.16\e{-4}$		& $-3.16\e{-4}$		\\
			&			& $-3.14\e{-4}$		& $-3.14\e{-4}$		\\
			&			&$-4\e{-6}$		& $-4\e{-6}$		\\
			\vspace{-3mm}\\\cline{2-4}\vspace{-3mm}\\
			& total		& $-6.35\e{-4}$		& $-6.35\e{-4}$		\\
			\\
			$6s6p\,\,^3\!P_0$
			& \PP		& $5.444\e{-3}$		& $-1.56\e{-5}$		\\
			&			& $2.53\e{-5}$		& $-7.24\e{-8}$		\\
			&			& $7.16\e{-6}$		& $-3.44\e{-6}$		\\
			&			& $1.9\e{-7}$		& $ 1.9\e{-7}$		\\
			&			& $-1.3\e{-8}$		& $-1.3\e{-8}$		\\
			\vspace{-3mm}\\
			& \D		& $-4.15\e{-4}$		& $-4.15\e{-4}$		\\
			&			& $-3.14\e{-4}$		& $-3.14\e{-4}$		\\
			&			& $-3\e{-6}$		& $-3\e{-6}$		\\
			\vspace{-3mm}\\\cline{2-4}\vspace{-3mm}\\
			& total		& $4.74\e{-3}$		& $-7.51\e{-4}$		\\
			\\
			differential
			& \PP		& $5.444\e{-3}$		& $-1.56\e{-5}$		\\
			&			& $2.53\e{-5}$		& $-7.24\e{-8}$		\\
			&			& $7.16\e{-6}$		& $-3.44\e{-6}$		\\
			&			& $0$				& $0$				\\
			&			&$-1.3\e{-8}$		& $-1.3\e{-8}$ 		\\
			\vspace{-3mm}\\
			& \D		& $-9.85 \e{-5}$	& $-9.85\e{-5}$		\\
			&			& $0$				& $0$				\\
			&			& $1\e{-6}$		& $1\e{-6}$      \\
			\vspace{-3mm}\\\cline{2-4}\vspace{-3mm}\\
			& total		& $5.379(10)\e{-3}$	& $-1.16(5)\e{-4}$	\\
			& expt.		& $5.363(6)\e{-3}$
		\end{tabular}
	\end{ruledtabular}
\end{table}
\subsection{The $^3\!P_0 -\, ^3\!P_1$ fine structure contribution}
In the dc limit, the $M1$ polarizability of the $6s6p\,^3\!P_0$ state is dominated by the $^3\!P_0 -\, ^3\!P_1$ fine structure contribution. This contribution is given by
\begin{gather*}
	\frac{2}{3}\frac{
		\left|\langle ^3\!P_0 ||\bm{\mu}||^3\!P_1 \rangle\right|^2}{h\nu_\mathrm{fs}},
\end{gather*}
where $\bm{\mu}$ is the magnetic dipole operator acting on the electrons and $\nu_\mathrm{fs}$ is the $^3\!P_0 -\, ^3\!P_1$ fine structure frequency splitting. We can partition the matrix element $\langle ^3\!P_0||\bm{\mu}||^3\!P_1 \rangle$ into a ``Dirac'' term and a ``Schwinger QED'' correction. The Dirac term assumes Dirac electrons, while the Schwinger QED term accounts for the electron's anomalous magnetic moment. In the nonrelativistic limit for electron motion, the terms combine to give~\cite{BreCheBel19}
\begin{gather*}
	\langle ^3\!P_0||\bm{\mu}||^3\!P_1 \rangle=\sqrt{2}\left(1+2a_e\right)\mu_B,
\end{gather*}
where $\mu_B$ is the Bohr magneton and $a_e$ is the electron magnetic moment anomaly. For the Dirac term, the relativistic CI+MBPT and CI+all order methods give results that are 0.85\% and 0.84\% below the nonrelativistic result, respectively. Meanwhile, relativistic calculations of the Schwinger QED term indicate that the nonrelativistic fractional correction $2a_e$ is sufficient at the level relevant for this work.

At the magic frequency, the $^3\!P_0 -\, ^3\!P_1$ fine structure contribution is scaled relative to the dc value by
\begin{gather*}
	\left[1-\left(\frac{\nu_L}{\nu_\mathrm{fs}}\right)^2\right]^{-1}
	\approx-\left(\frac{\nu_\mathrm{fs}}{\nu_L}\right)^{2}
	\approx-\frac{1}{285},
\end{gather*}
where $\nu_L$ is the lattice frequency. Meanwhile, the diamagnetic contribution is insensitive to the drive frequency (dc or magic). It follows that, while the $^3\!P_0 -\,^3\!P_1$ fine structure contribution is dominant and the diamagnetic contribution is subdominant in the dc limit, the opposite holds true at the magic frequency.
\subsection{Experimental value for the differential $M1$ polarizability in the dc limit}
 
The differential $M1$ polarizability in the dc limit, when multiplied by a factor of $-1/2h$, simply equates to the quadratic Zeeman shift (QZS) coefficient, which quantifies the frequency shift to the clock transition in the presence of a dc magnetic field. Reference~\cite{mcgrew2018atomic} reports a QZS coefficient $-0.06095(7)~\text{Hz}/\text{G}^2$, using the experimental method outlined in Ref.~\cite{BoyZelLud07}. To evaluate the QZS coefficient, the dc magnetic field was calibrated using the $M1$ moment of the neutral $^{171}$Yb system in the ground electronic state ($F=1/2$). This atomic $M1$ moment was measured in Ref.~\cite{Ols72}. By applying a theoretical diamagnetic correction factor on the order of a percent (an effect distinct from the diamagnetic shift discussed elsewhere in this work), the $M1$ moment of the bare $^{171}$Yb nuclear system ($I=1/2$) can be inferred from the atomic $M1$ moment~\cite{FeiJoh68,Ful76}. Both the measured atomic $M1$ moment and the inferred nuclear $M1$ moment are reported in~\cite{Ols72}. After inspecting laboratory notes, we have found that the wrong $M1$ moment was applied in the evaluation of the QZS coefficient. Appropriately rescaling the value in~\cite{mcgrew2018atomic}, here we report an updated QZS coefficient of $\left[-0.06095(7)~\text{Hz}/\text{G}^2\right]\left(0.487937/0.491889\right)^2=-0.05997(7)~\text{Hz}/\text{G}^2$. We clarify that the theoretical diamagnetic correction factor, while needed to infer the nuclear $M1$ moment from the atomic $M1$ moment, ultimately does not play a role in the determination of the QZS coefficient. Using this updated value for the QZS coefficient, we obtain a result of $5.363(6)\times10^{-3}$~a.u.\ for the differential $M1$ polarizability in the dc limit. Here a.u.\ denotes atomic units based on Gaussian electromagnetic expressions (e.g., the Bohr magneton has a value of $\alpha/2$ in these units, where $\alpha$ is the fine structure constant).

We note that no clock error is attributed to the erroneous QZS coefficient. Operationally, the QZS is determined from $m_F$-dependent line splittings of the clock transition. The corresponding QZS coefficient [i.e., (clock frequency shift)/(frequency splitting)$^2$] was measured and subsequently applied; error only occurred in the conversion of this coefficient to absolute units of magnetic field [(clock frequency shift)/(magnitude of magnetic field)$^2$].

\subsection{Final $\tilde{\alpha}_{M1E2}$ value}
To obtain the theoretical value for the parameter $\tilde{\alpha}_{M1E2}$ presented in the main text, we divide the differential $M1$+$E2$ polarizability at the magic frequency ($-9.0(2.4)\e{-5}$~a.u.) by the $E1$ polarizability at the magic frequency (common to both clock states by definition; 186 a.u.~\cite{DzuDer10}) and multiply this ratio by the ratio of the lattice recoil energy to the clock photon energy ($3.88\e{-12}$). This yields $\tilde{\alpha}_{M1E2}= -1.9(5)\e{-18}$.
\begin{table}
	\caption{Computed $E2$ polarizabilities for the Yb clock states in the CI+all-order approximation at the magic frequency.
		All values are in atomic units.}
	\label{Tab:E2}
	\begin{ruledtabular}
		\begin{tabular}{llc}
			&		 		& magic frequency	\\
			\hline
			\vspace{-3mm}\\
			$6s^2\,^1S_0$	& valence		& $5.0\e{-5}$		\\
			& core			& $1.8\e{-7}$		\\
			& valence-core	& $-2\e{-12}$	\\
			\cline{2-3}\vspace{-3mm}\\
			& total			& $5.0\e{-5}$		\\
			\\
			$6s6p\,^3P_0$	& valence		& $7.6\e{-5}$		\\
			& core			& $1.8\e{-7}$		\\
			& valence-core	& $-6\e{-9}$		\\
			\cline{2-3}\vspace{-3mm}\\
			& total			& $7.6 \e{-5}$		\\
			\\
			differential	& valence		& $2.6\e{-5}$		\\
			& core			& $0$				\\
			& valence-core	& $-6\e{-9}$		\\
			\cline{2-3}\vspace{-3mm}\\
			& total			& $2.6\e{-5}$
		\end{tabular}
	\end{ruledtabular}
\end{table}

\begin{table}
	\caption{Computed $M1$, $E2$, and $M1$+$E2$ polarizabilities for the Yb clock states at the magic frequency in
		the CI+MBPT and CI+all-order approximations. All values are in atomic units based on Gaussian electromagnetic expressions.}
	\label{Tab:combined}
	\begin{ruledtabular}
		\begin{tabular}{lcrr}
			&			&   CI + MBPT       &  CI + all-order \\
			\hline
			\vspace{-3mm}\\
			$6s^2\,^1S_0$	& $M1$		& $-6.32\e{-4}$	    & $-6.35\e{-4}$	   \\
			& $E2$		& $ 0.48\e{-4}$	    & $ 0.50(7)\e{-4}$	   \\
			\cline{2-4}\vspace{-3mm}\\
			& $M1+E2$	& $-5.84\e{-4}$	    & $-5.84\e{-4}$	   \\
			\\
			$6s6p\,^3P_0$	& $M1$		& $-7.53\e{-4}$	    & $-7.51\e{-4}$	   \\
			& $E2$		& $ 0.79\e{-4}$	    & $ 0.76(23)\e{-4}$	   \\
			\cline{2-4}\vspace{-3mm}\\
			& $M1+E2$	& $-6.74\e{-4}$	    & $-6.75\e{-4}$	   \\
			\\
			differential	& $M1$		& $-1.21\e{-4}$	    & $-1.16(5)\e{-4}$	\\
			& $E2$		& $ 0.31\e{-4}$	    & $ 0.26(23)\e{-4}$\\
			\cline{2-4}\vspace{-3mm}\\
			& $M1+E2$	& $-0.90\e{-4}$     & $-0.90(24)\e{-4}$
		\end{tabular}
	\end{ruledtabular}
\end{table}

\bibliography{references.bib}